\documentclass[aps,prl,notitlepage,tightenlines,nofootinbib,superscriptaddress,twocolumn]{revtex4-2}

\usepackage{multirow}

\usepackage{amsmath}
\usepackage{amssymb,amsthm}
\usepackage{mathtools}
\usepackage{mathrsfs}
\usepackage{relsize}
\usepackage{bm}
\allowdisplaybreaks[4]  
\usepackage{epsfig,graphics,graphicx}
\usepackage{color,xcolor}
\usepackage{subfigure}
\usepackage{array}
\usepackage{ragged2e}
\usepackage{lineno}
\usepackage{natbib}
\usepackage{hyperref}
\hypersetup{colorlinks=true,
            linkcolor=blue,
            anchorcolor=blue,
            citecolor=darkgreen,
            filecolor=blue,
            urlcolor=blue,
            bookmarksnumbered=true,
            pdfview=FitB
}
\usepackage{enumitem}
\usepackage{ulem}
\usepackage{soul}
\usepackage{listings}
\colorlet{darkgreen}{green!50!black}
\colorlet{brightyellow}{yellow!75!red}
\colorlet{orange}{red!50!yellow}
\colorlet{darkgray}{gray!50!black}
\colorlet{darkred}{red!50!black}
\def\dd{{\mathrm{d}}}

\newcommand{\half}[1][1] {\mathsmaller{\frac{#1}{2}}}

\makeatletter
\newcommand*{\transpose}{%
  {\mathpalette\@transpose{}}%
}
\newcommand*{\@transpose}[2]{%
  \raisebox{\depth}{$\m@th#1\intercal$}%
}
\makeatother 
\begin{document}

\title{Stress inside the pion in holographic light-front QCD}

\author{Yang Li}
\affiliation{University of Science and Technology of China, Hefei, Anhui 230026, China}

\author{James P. Vary}
\affiliation{Iowa State University, Ames, IA 50011, USA}

\date{\today}

\begin{abstract}
In this work, we propose a method to compute the gravitational form factor $D(Q^2)$ in holographic QCD by exploiting the remarkable correspondence between semi-classical light-front QCD and semi-classical field theories in wrapped spacetime in 5D. The use of light-front holography bridges physics at large $Q^2$ as attained in light-front QCD and physics at small $Q^2$ where the coupling to scalar and tensor fields, e.g. glueballs, are dominant. As an application, we compute the $D$-term for the pion, and compare the results with recent lattice simulations. 
\end{abstract}
\maketitle

\paragraph{Introduction}
The hadronic energy-momentum tensor (EMT) characterizes the energy, momentum and stress distributions inside the composite systems of quarks held together by the strong force. It also controls how the hadronic matter -- 99\% of all visible matter -- gravitates. Due to the strong-coupling nature of QCD, a thorough understanding of the hadronic EMT is still missing \cite{Gross:2022hyw}. Indeed, one of the global charges associated with the hadronic EMT, the $D$-term, is dubbed as the ``last global unknown" \cite{Polyakov:2018zvc}. Unlike the mass and spin, this term is not constrained by global conservations laws. It encodes dynamical information of the system, such as the force balance and mechanical stability. The $D$-term has recently attracted attention from both the theory and experimental communities \cite{Burkert:2023wzr}. 

Light-front QCD (LFQCD) provides a crucial nonperturbative picture of the hadronic EMT. The gravitational form factors (GFFs) are related to the second Mellin moments of the generalized parton distributions (GPDs), which also allows a further decomposition of the total energy, spin and stress distributions into quark, gluon and anomalous contributions \cite{Ji:1997gm, Polyakov:2002yz}. The light-front wave function (LFWF) representation also affords a direct microscopic interpretation of the hadronic EMT \cite{Brodsky:2000ii, Cao:2023ohj}. The LFWFs can be obtained by diagonalizing the QCD Hamiltonian quantized at fixed light-front time $x^+ = x^0 + x^3$ as demonstrated by the basis light-front quantization approach (BLFQ \cite{Vary:2009gt}). Recent strides in this direction include successful description of quarkonium spectra \cite{Li:2015zda, Li:2017mlw, Tang:2018myz, Tang:2019gvn, Qian:2020utg, Lan:2021wok}, radiative transitions \cite{Li:2018uif, Li:2021ejv}, form factors \cite{Adhikari:2018umb, Mondal:2019jdg}, nucleon valence, sea and gluon parton distributions (PDFs, \cite{Lan:2019vui, Xu:2021wwj, Adhikari:2021jrh, Liu:2022fvl, Xu:2022abw, Lin:2023ezw, Xu:2023nqv}), and transverse momentum distributions \cite{Hu:2022ctr, Zhu:2023lst, Zhu:2023nhl}. 

One of the most remarkable features of light-front QCD is its connection to the holographic view of QCD \cite{Brodsky:2003px} (see Ref.~\cite{Brodsky:2014yha} for a review). Holographic QCD or AdS/QCD is inspired by the AdS/CFT correspondence, a duality relating strongly coupled gauge theory with weakly interacting gravity theory, and paves the way to access the strongly coupled regime of QCD while retaining the analytic power \cite{Maldacena:1997re, Polchinski:2000uf, Polchinski:2001tt, Erlich:2005qh}. 

One of the leading examples is soft-wall AdS/QCD \cite{Karch:2006pv}. This model introduces a scale through a dilaton field $\phi(z) = \kappa^2 z^2$ to break the conformal symmetry to match QCD. The scale parameter $\kappa = 0.388\,\mathrm{GeV}$ is obtained by fitting to the $\rho$ mass and has been adopted to investigate form factors in AdS/QCD \cite{Erlich:2005qh, Karch:2006pv, Grigoryan:2007my, Kwee:2007dd, Brodsky:2007hb, Abidin:2008ku, Abidin:2009hr, Abidin:2010rlp}. 
The hadron mass spectra successfully reproduce the Regge trajectory, $M^2_n = \alpha n + \beta$. 
Dynamical information of the hadrons such as the decay constant, electromagnetic form factors and transition form factors  was also computed and compared with experiments with reasonable agreement \cite{Grigoryan:2007vg, Grigoryan:2007my, Kwee:2007dd, Grigoryan:2007wn, Abidin:2008ku, Grigoryan:2008up, Abidin:2008hn, Abidin:2009hr, Abidin:2009aj, Grigoryan:2009pp, Abidin:2019xwu}. 
The fifth dimension $z$ of the AdS space in holographic QCD is shown to be associated with the parton impact parameter $\zeta_\perp = \sqrt{x(1-x)}r_\perp$, where $x$ is the longitudinal momentum fraction of the parton and $r_\perp$ is the transverse separation of the partons. This striking correspondence bridges the holographic view with the partonic picture and provides an avenue to compute PDFs as well as GPDs in holographic QCD \cite{deTeramond:2018ecg, Liu:2019vsn, deTeramond:2021lxc, Dosch:2022mop}. 

In this note, we compute the GFF $D_\pi(Q^2)$ of the pion in holographic light-front QCD (HLFQCD). This form factor emerges from the covariant decomposition of the pionic matrix element of the EMT operator \cite{Kobzarev:1962wt}, 
\begin{multline}
\langle p+q | T^{\mu\nu}(0) | p \rangle \\
= 2P^\mu P^\mu A_\pi(-q^2) + \frac{1}{2}(q^{\mu}q^\nu - q^2 g^{\mu\nu}) D_\pi(-q^2)
\end{multline} 
where $q = p'-p, P = (p'+p)/2$. The GFFs $A_\pi$ and $D_\pi$ can be obtained from two EMT components within the Drell-Yan frame $q^+=0, P_\perp = 0$, 
\begin{align}
 \langle p+q | T^{++} | p \rangle =& 2P^+ P^+ A_\pi(Q^2), \label{eqn:T++} \\
 \langle p+q | T^{+-} | p \rangle =& (2 M^2_\pi+ \half Q^2) A_\pi(Q^2) +  Q^2 D_\pi(Q^2) \label{eqn:T+-}
\end{align} 
Here we adopt light-front coordinates $v^\pm = v^0 \pm v^3$ and $\vec v_\perp = (v^1, v^2)$ for a 4-vector $v^\mu$ \cite{Gross:2022hyw, Brodsky:1997de}, and $Q^2 =-q^2 = q_\perp^2$. Since  $x^+$ in light-front dynamics is the ``time", $T^{+\mu}$ is the conserved Noether current associated with the conserved 4-momentum \cite{Brodsky:1997de}, 
\begin{equation}
P^\mu = \frac{1}{2}\int  \dd x^-\dd^2x_\perp\, T^{+\mu}(x).
\end{equation}
The conservation of 4-momentum implies \cite{Polyakov:2018zvc} 
\begin{equation}\label{eqn:von_Laue_condition}
A_\pi(0) = 1, \quad \lim_{Q\to0} Q^2D_\pi(Q^2) = 0.
\end{equation} 
The second expression is known as von Laue condition \cite{Polyakov:2018zvc}, which suggests the force balance inside the pion. It was further conjectured that the mechanically stable system has a negative $D$-term $D = D(0) < 0$ \cite{Polyakov:2018zvc}. For the pion, chiral perturbation theory predicts that in the chiral limit $D_\pi = -1$, the same as free scalar particles \cite{Donoghue:1991qv}.

\begin{figure}
\;\;\;\includegraphics[width=0.43\textwidth]{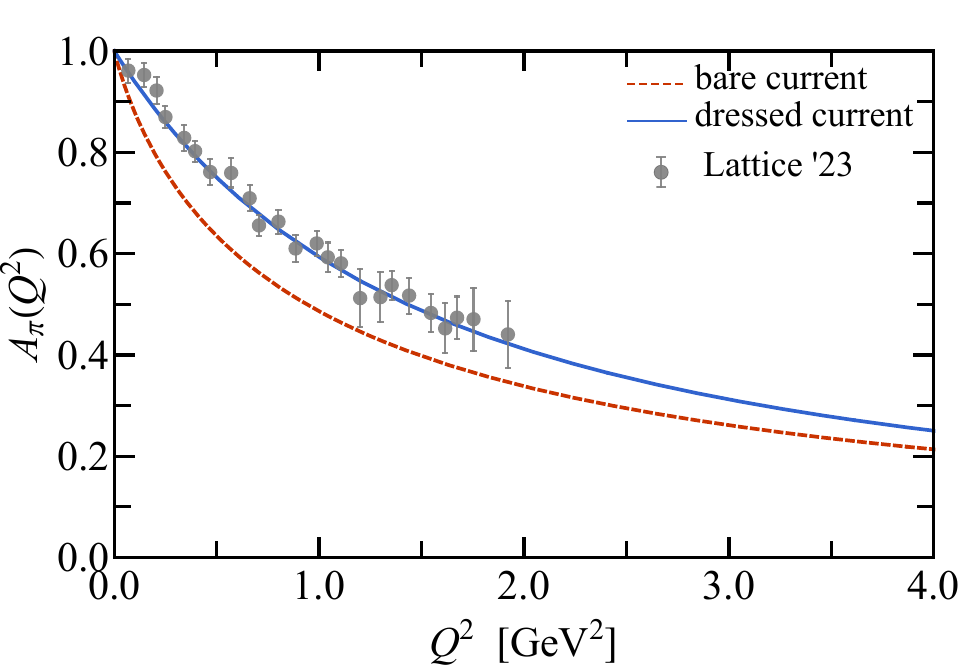}
\includegraphics[width=0.45\textwidth]{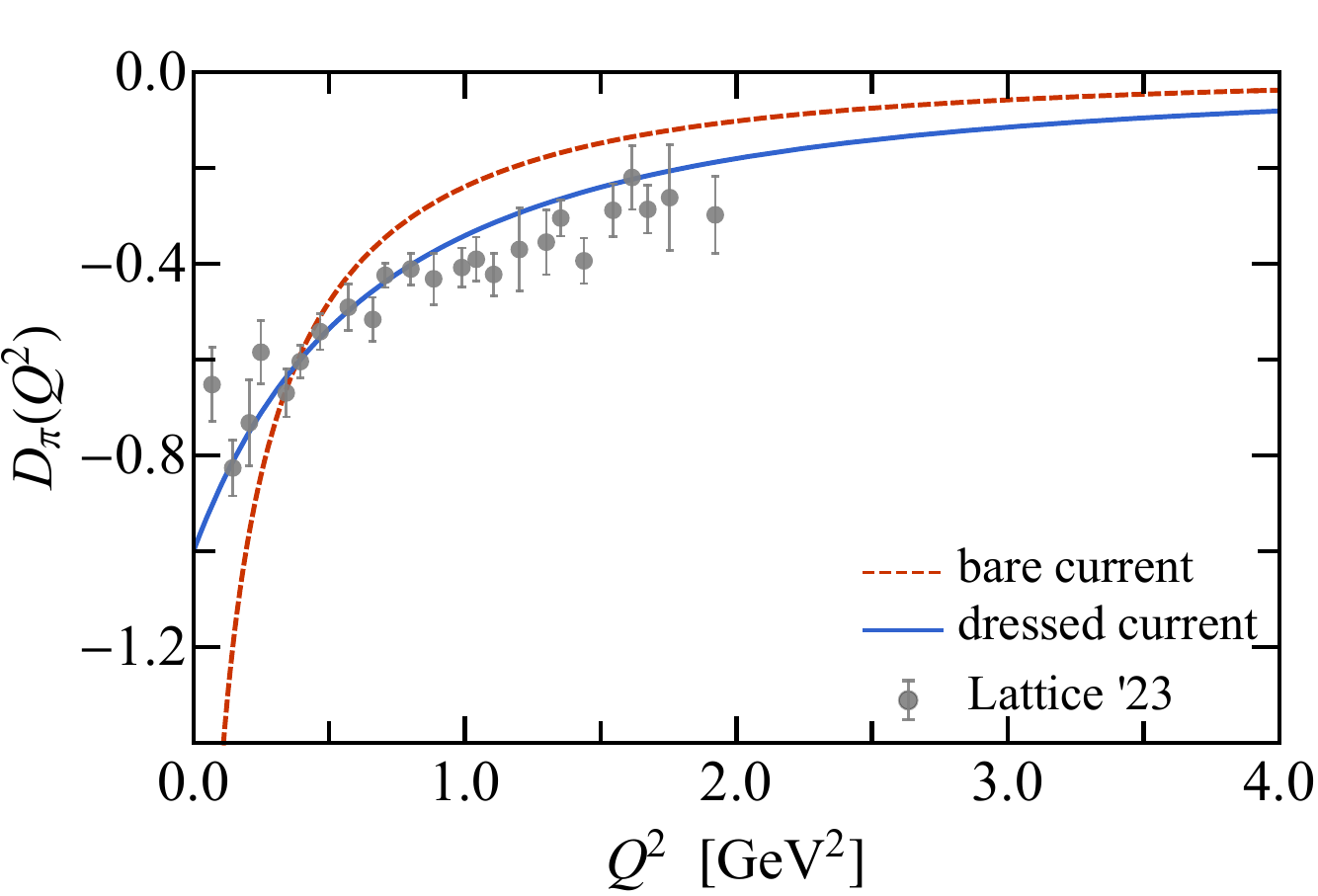}
\caption{Pion GFFs from HLFQCD as compared with the recent Lattice QCD results with a pion mass 170 MeV \cite{Hackett:2023nkr}. (Top) $A_\pi(Q^2)$; (Bottom) $D_\pi(Q^2)$.  The soft-wall scale parameter is $\kappa = 0.388\,\mathrm{GeV}$.}
\label{fig:GFF}
\end{figure}

\paragraph{Gravitational form factor $A_\pi$ in light-front holography}
The GFF $A_\pi(Q^2)$ can be represented by the LFWFs as \cite{Brodsky:2000ii},
\begin{equation}\label{eqn:A_coord}
A_\pi(q_\perp^2) =\sum_n \int \big[\dd x_i \dd^2 r_{i\perp}\big]_n  \big| \psi_n(\{x_i, \vec r_{i\perp}\})\big|^2 \sum_j x_j e^{i \vec r_{j\perp}\cdot \vec q_\perp}
\end{equation}
where $\psi_n(\{x_i, \vec r_{i\perp}\})$ is the $n$-body LFWF, and $\big[\dd x_i \dd^2 r_{i\perp} \big]_n$ the $n$-body phase space measure.
Using soft-wall LFWFs of the valence Fock sector ($q\bar q$) \cite{Brodsky:2006uqa},
\begin{equation}\label{eqn:holographic_LFWF}
\psi_\pi(x, \vec r_\perp) = \sqrt{4\pi}x(1-x) \varphi_\pi(\zeta_\perp) 
\end{equation}
where $\varphi_\pi(\zeta_\perp) = ({\kappa}/{\sqrt{\pi}})\exp\big\{-\frac{1}{2}\kappa^2\zeta_\perp^2\big\}$ in the soft-wall model, the valence contribution can be written explicitly as, \cite{Brodsky:2008pf}
\begin{equation}\label{eqn:Api}
A_\pi(Q^2) = \int \dd^2 \zeta_\perp \, \big| \varphi_\pi(\zeta_\perp) \big|^2 \frac{1}{2}Q^2 \zeta_\perp^2 K_2(\zeta_\perp Q) + \cdots
\end{equation} 
The ellipses indicate higher Fock sector contributions, which are important at low  resolution $Q \lesssim \Lambda_\textsc{qcd}$. Comparing this expression with GFF $A_\pi$ obtained directly in AdS/QCD \cite{Abidin:2008ku, Abidin:2008hn}, one identifies  $(1/2)Q^2 \zeta_\perp^2 K_2(\zeta_\perp Q)$ as the UV part of the dressed current, 
$H(Q^2, z) = (1/2)z^2 Q^2 K_2(z Q) + \cdots $ \cite{Brodsky:2008pf}. Thus, the high Fock sector contributions can be absorbed into the dressed current,
\begin{equation}
A_\pi(Q^2) = \int \dd^2 \zeta_\perp \, \big| \varphi^2_\pi(\zeta_\perp) \big|^2 H(Q^2, \zeta_\perp)
\end{equation} 
In the soft-wall model, this current is attained from the bulk-to-boundary propagator of a tensor field \cite{Abidin:2009hr},
\begin{equation}
H(Q^2, z) = \Gamma\Big(2+\frac{a}{2}\Big) U\Big(\frac{a}{2}, -1; 2\xi\Big) 
\end{equation}
where, $a = Q^2/(4\kappa^2), \xi = \kappa^2z^2$, $\Gamma(z)$ is Euler's Gamma function, and $U(a, b; z)$ is the second Kummer function \cite{DLMF}. These results, originally obtained in Ref.~\cite{Abidin:2009hr} for soft-wall AdS/QCD and in Ref.~\cite{Brodsky:2008pf} for HLFQCD, are reproduced in Fig.~\ref{fig:GFF} in comparison with the recent Lattice prediction \cite{Hackett:2023nkr}. 

\paragraph{Gravitational form factor $D_\pi$ in light-front holography}

In contrast to $T^{++}$ (hence GFF $A_\pi$), $T^{+-}$ is the conserved current of the light-front Hamiltonian $P^-$, which contains interactions. To preserve the von Laue condition (\ref{eqn:von_Laue_condition}), a consistent treatment of the interaction is required. Ref.~\cite{Chakrabarti:2020kdc} computed the nucleon $D(Q^2)$ by adopting the free EMT operator with an effective LFWF that corresponds to the dressed vector current $J^\mu$ in soft-wall AdS/QCD. The obtained GFF $D(Q^2)$ does not satisfy the von Laue condition (\ref{eqn:von_Laue_condition}). 

On the AdS/QCD side, GFFs $A(Q^2)$ can be extracted by coupling the hadronic fields to a transverse and traceless gravitational field in 5D AdS space \cite{Abidin:2008ku, Abidin:2008hn, Abidin:2009hr}.  Unfortunately, the same procedure does not access the $D$-term. In a series of recent works~\cite{Mamo:2019mka, Mamo:2021tzd, Mamo:2022eui}, the authors found that the nucleon GFF $D(Q^2)$ vanishes in the leading order of $O(1/N_c)$, and was proportional to GFF $A(Q^2)$ in the next to leading order with the proportionality $D(0)$ undetermined. Ref.~\cite{Fujita:2022jus} computed the nucleon $D(Q^2)$ in the Sakai-Sugimoto model \cite{Sakai:2004cn, Sakai:2005yt}, a top-down model in 10D whose low-energy dual closely resembles QCD. The authors observed contributions from scalar ($0^{++}$) and tensor ($2^{++}$) glueballs. 

Here, we start from the QCD side with a LFWF representation. This approach is closely related to the pQCD analysis \cite{Tong:2021ctu, Tong:2022zax}. We split the EMT operator into the kinetic energy part and potential energy (interaction) part $T^{+-} = T^{+-}_0 + T^{+-}_\text{int}$, which corresponds to Hamiltonians,
\begin{equation}
P^-_0 = \int \dd^3x\, T^{+-}_0(x),  \quad
P^-_\text{int} = \int \dd^3x\, T^{+-}_\text{int}(x).
\end{equation}
The pionic matrix element of $T^{+-}_0$ admits an exact LFWF representation as \cite{Cao:2023ohj}, 
\begin{multline}
T(Q^2) \equiv \langle p+q|T^{+-}_0|p\rangle = \int \big[\dd x_i \dd^2 r_{i\perp} \big]_n \psi_n^*(\{x_i, \vec r_{i\perp}\}) \\
\times \sum_j  e^{i\vec r_{j\perp}\cdot\vec q_\perp} \frac{-\nabla_{j\perp}^2+m_j^2-\frac{1}{4}q_\perp^2}{x_j} 
 \psi_n(\{x_i, \vec r_{i\perp}\}) 
\end{multline}

For the interaction part $V(Q^2) \equiv \langle p+q|T^{+-}_\text{int}|p\rangle$, energy conservation implies 
\begin{equation}\label{eqn:energy_conservation}
T(0) + V(0) = 2M^2_\pi = 0
\end{equation}
This expression is equivalent to the von Laue condition (\ref{eqn:von_Laue_condition}). To preserve  (\ref{eqn:von_Laue_condition}) or (\ref{eqn:energy_conservation}), it is necessary to adopt an operator $T^{+-}_\text{int}$ in accordance with the hadronic LFWFs. 
For the soft-wall LFWF, the corresponding light-front potential energy operator $P^-_\text{int}$ is attained by second quantizing the soft-wall potential
$U_\textsc{sw}(x, r_\perp) = \kappa^2(\kappa^2 \zeta^2_\perp - 2)$ where $\zeta_\perp = \sqrt{x(1-x)}r_\perp$. $T^{+-}_\text{int}(x)$ is its local density operator, which can be obtained by localizing the light-front quantized operator $P^-_\text{int}$ in the transverse direction. 
Operator localization on the light front is similar to that in non-relativistic quantum many-body theory. For example, the transverse charge density can be obtained by inserting the 2D Dirac-$\delta$ in the many-body formulation, i.e.,
$Q = \sum_i e_i \; \to \;  \rho(r_\perp) = \sum_i e_i \delta^2(r_\perp - r_{i\perp})$. By the same token, the soft-wall interaction can be written in a many-body form,
$H_\text{sw} = \frac{1}{2!}\sum_{i,j} U_\text{sw}(x_{ij}, \vec r_{i\perp}-\vec r_{j\perp})$
where $x_{ij} = {x_ix_j}/{(x_i+x_j)}$ and $x_i = p^+_i/P^+$ is the longitudinal momentum fraction of the $i$-th parton, $\vec r_{i\perp}$ is the transverse coordinate of the $i$-th parton. Note that this is a two-body operator and can be localized by inserting Dirac-$\delta$'s in the transverse direction:
\begin{multline}
\widetilde V(r_\perp) = \frac{1}{2!}\sum_{i,j} U_\text{sw}(x_{ij}, \vec r_{i\perp}-\vec r_{j\perp}) \\
\times \frac{1}{2}\Big\{\delta^2(r_\perp - r_{i\perp}) + \delta^2(r_\perp - r_{j\perp}) \Big\},
\end{multline}
which can be converted to the wave function representation following the standard second quantization procedure.
%
The resulting potential energy density in the valence sector is, 
\begin{multline}
\widetilde V(x_\perp) =  \frac{1}{2}\int \frac{\dd x}{4\pi x(1-x)} \int \dd^2 r_{\perp} \big|\psi(x, r_\perp)\big|^2 U_\textsc{sw}(x, r_\perp) \\
\times 
\Big\{\delta^2\big(x_\perp - (1-x)r_\perp\big) + \delta^2(x_\perp + xr_\perp)\Big\}
\end{multline}
Plugging in the soft-wall wave function (\ref{eqn:holographic_LFWF}), the pion GFF $D_\pi(Q^2)$ reads, 
\begin{multline}\label{eqn:D_pi_bare}
D_\pi(Q^2) 
= \int \dd^2 \zeta_\perp  \big|\varphi _\pi(\zeta_\perp)\big|^2 
 \Big\{  
 \frac{\zeta_\perp^2 Q^2 }{4} K_2(\zeta_\perp Q) - 2K_0(\zeta_\perp Q)  \\
- \frac{2U_\textsc{sw}(\zeta_\perp)}{Q^2}  \Big[ \zeta_\perp Q K_1(\zeta_\perp Q) - \frac{1}{2} \zeta^2_\perp Q^2 K_2(\zeta_\perp Q) \Big]  \Big\}
\end{multline}
where $K_\nu(z)$ is the modified Bessel function of the second kind \cite{DLMF}, $U_\textsc{sw}(\zeta_\perp) = \kappa^2(\kappa^2\zeta_\perp-2)$ is the (soft-wall) holographic potential. It is not hard to see this expression satisfies the von Laue condition (\ref{eqn:von_Laue_condition}) as expected. 
Eq.~(\ref{eqn:D_pi_bare}) is the main result in this note. Fig.~\ref{fig:GFF} shows Eq.~\ref{eqn:D_pi_bare} (bare current) in comparison with the recent Lattice calculation \cite{Hackett:2023nkr}. 
At large $Q^2$, the form factor scales as $1/Q^2$, in agreement with the pQCD analysis \cite{Tong:2022zax}. 
 
Similar to Eq.~(\ref{eqn:Api}), the valence ansatz captures physics at large $Q^2$. In the forward limit $Q^2 \to 0$, the obtained $D_\pi(Q^2)$ diverges. 
To better incorporate physics beyond the valence Fock sector, we can dress the current. One readily identifies $\zeta_\perp Q K_1(\zeta_\perp Q)$ and $\frac{1}{2}\zeta_\perp^2Q^2K_2(\zeta_\perp Q)$ as the UV parts of the dressed vector and tensor currents \cite{Brodsky:2014yha}
\begin{align}
V(Q^2, \zeta_\perp) =  \Gamma(1+a) U(a, 0; \xi), \\
H(Q^2, \zeta_\perp) = \Gamma(2+\frac{a}{2}) U(\frac{a}{2}, -1; 2\xi),
\end{align} 
respectively with $a = Q^2/(4\kappa^2)$ and $\xi = \kappa^2\zeta_\perp^2$. 
$K_0(\zeta_\perp Q)$ may be identified as the bare scalar current. In principle, the dressing of this current can be obtained from the bulk-to-boundary propagator of the scalar fields. 
The IR finite bulk-to-boundary propagators of the scalar field with conformal dimension $\Delta = 3$ (i.e. scalar mesons) and 
$\Delta = 4$ (i.e. scalar glueballs) in soft-wall model are $S_{\Delta=3} = z\Gamma(a+\frac{3}{2}) U(a+\frac{1}{2}, 0, \xi)$ and $S_{\Delta=4} = \Gamma(a+2)U(a, -1, \xi)$, respectively \cite{Colangelo:2007pt, Forkel:2007ru, Colangelo:2008us}. We assume both currents contribute to the $D$-term, 
\begin{equation}
S(Q^2, \zeta_\perp) = c_1 S_{\Delta=3}(Q^2, \zeta_\perp) + c_2 S_{\Delta=4}(Q^2, \zeta_\perp)
\end{equation}
and their coefficients $c_1 = 0.79, c_2 = 0.25$ are determined from matching to both the UV asymptotics (\ref{eqn:D_pi_bare}) and the IR limit $D_\pi(0) = -1$ as predicted by the chiral perturbation theory \cite{Donoghue:1991qv}. 
The resulting GFF $D_\pi(Q^2)$ with the dressed currents is, 
\begin{multline}\label{eqn:D_pi_dressed}
D_\pi(Q^2) 
= \int \dd^2 \zeta_\perp \big| \varphi_\pi(\zeta_\perp) \big|^2 
 \Big\{  
 \frac{1}{2} H(Q^2, \zeta_\perp) - 2S(Q^2, \zeta_\perp)  \\
  - \frac{2U_\textsc{sw}(\zeta_\perp)}{Q^2} \Big[ V(Q^2, \zeta_\perp) - H(Q^2, \zeta_\perp) \Big] \Big\}.
\end{multline}
Fig.~\ref{fig:GFF} compares results with the dressed current (\ref{eqn:D_pi_dressed}) and the bare current (\ref{eqn:D_pi_bare}), together with the recent Lattice calculation \cite{Hackett:2023nkr} (cf. Ref.~\cite{Xu:2023izo}).  
The dressed current are in good agreement with with recent Lattice simulation at $Q^2\lesssim 1\,\mathrm{GeV}^2$. 
From these results, we can extract the mechanical radius of the pion, $r_\text{mech}^2 \equiv -6 D'(0)/D(0) = (0.60\,\mathrm{fm})^2$, which is larger than the predicted pion matter (mass) radius $r^2_\text{mat} \equiv -6 A'(0)/A(0) = (0.39\,\mathrm{fm})^2$. 
These values are very close to the corresponding radii $r^2_\text{mat} = (0.41\pm0.01\,\mathrm{fm})^2$ and $r_\text{mech}^2 = (0.61\pm0.07\,\mathrm{fm})^2$ extracted from the recent Lattice simulation \cite{Hackett:2023nkr}, while differ from the mechanical radius $r_\text{mech}^2 = (0.82-0.88\,\mathrm{fm})^2$ extracted from the two photon process $\gamma^*\gamma \to \pi^0 \pi^0$ by $\sim$30\% \cite{Kumano:2017lhr}. 

\paragraph{Summary and discussions} 

In this note, we investigated the pion gravitational form factors by exploiting the remarkable correspondence between light-front QCD and gravity theory in 5D. We computed the $D$-term of the pion using the bare EMT on the light front valid at high resolution. To extend the results to low resolution probes, we adopted scalar, vector and tensor dressed currents based on bulk-to-boundary propagators for scalar, vector and tensor fields. 

The $D$-term of the pion stems from several sources. The largest contribution stems from the coupling to the scalar mesons, $D_\pi^M(0) = -0.83$. 
This current contains a series of scalar meson poles $M_n^2 = (4n+6)\kappa^2$ in the timelike region \cite{Colangelo:2008us}, 
\begin{equation}
D_\pi^M(-q^2) = \sum_{n=0} \frac{f_n}{q^2-M_n^2}.
\end{equation}
The lowest scalar meson mass $M = \sqrt{6}\kappa = 0.95 \,\mathrm{GeV}$ is close to the mass of $a_0(980)$. 
The scalar glueball contributes to a pole term 
\begin{equation}
D_\pi^G(-q^2) = \frac{4\kappa^2}{q^2 - 8\kappa^2}
\end{equation}
The pole mass $M_G = \sqrt{8}\kappa = 1.1\,\mathrm{GeV}$ is the mass of the ground-state scalar glueball predicted in AdS/QCD. The corresponding 
$D$-term is $D_\pi^G(0) = -1/2$. 
The contribution of the tensor current has an opposite sign, $D_\pi^T(0) = +1/2$. 

The contribution of the last term in (\ref{eqn:D_pi_dressed}) is small, $D_\pi^{V-T}(0) = -0.17$. 
Chiral symmetry plays a pivotal role here. In the hard-wall model, the corresponding term vanishes completely. For other hadrons, the emergent hadron mass is expected to give a large contribution that significantly alters the budget of the $D$-term.  
This term consists of a vector current and a tensor current. However, in contrast to the nucleon case, the $\rho$-pole at $M^2_\rho = 4\kappa^2$ does not contribute to the $D$-term. 
In light-front dynamics, according to pQCD dimensional counting rules \cite{Brodsky:1973kr, Brodsky:1974vy}, the vector current contains a twist-3 ($|q\bar q g\rangle$) contribution and a twist-2 contribution ($|q\bar q\rangle$). It is interesting to see the net result is the twist-3 contribution minus the twist-2 result, leaving the pure gluon contribution. This cancellation relies critically on the fact that the pion is massless in the chiral limit. 
Its remarkable that the particle identification in AdS/QCD is consistent with the pQCD counting rule \cite{Brodsky:1973kr, Brodsky:1974vy}. 

We note that the picture presented here is qualitatively similar to the recent holographic QCD computation of the nucleon GFF $D(Q^2)$ within the Sakai-Sugimoto model in 10 dimensions \cite{Fujita:2022jus}. 
Furthermore, the method we proposed here can also be extended to the nucleon sector and be compared with the results from these models directly. 

Note that we are considering only the chiral limit. Refs.~\cite{Li:2021jqb, deTeramond:2021yyi, Li:2022izo} showed the longitudinal degrees of freedom need to be incorporated for finite quark masses. Then how to establish the correspondence between AdS/QCD and light-front dynamics is an interesting question to explore in the next step. 

\paragraph{Acknowledgements}
The authors acknowledge fruitful discussions with Stan Brodsky and Guy de Téramond. 
Y.L. is supported by the New faculty start-up fund of the University of Science and Technology of China. 
 This work was supported in part by the National Natural Science Foundation of China (NSFC) under Grant No.~12375081, by the Chinese Academy of Sciences under Grant No.~YSBR-101, and by the US Department of Energy (DOE) under Grant No. DE-SC0023692.


\end{document}